\documentclass[conference,table]{IEEEtran}
\IEEEoverridecommandlockouts
\usepackage[utf8]{inputenc}
\usepackage{cite}
\usepackage{amsmath,amssymb,amsfonts}
\usepackage{algorithmic}
\usepackage{graphicx}
\usepackage{textcomp}
\usepackage[os=win]{menukeys}
\usepackage{cleveref}
\usepackage{listings}
\usepackage{lipsum}
\usepackage{wrapfig}
\usepackage{rotating}
\usepackage{lscape}
\usepackage{tabularx}
\usepackage{tikz}
\usepackage{verbatim}
\usepackage{times}    
\usepackage{multirow,booktabs,array,tabulary,longtable,xltabular,caption}
\usepackage{newtxtext,newtxmath}

\def\BibTeX{{\rm B\kern-.05em{\sc i\kern-.025em b}\kern-.08em
    T\kern-.1667em\lower.7ex\hbox{E}\kern-.125emX}}
\begin{document}

\title{A Threat Intelligence Event Extraction Conceptual Model for Cyber Threat Intelligence Feeds\\
}
\author{\IEEEauthorblockN{1\textsuperscript{st}Jamal H. Al-Yasiri, Member, IEEE}
\IEEEauthorblockA{\textit{Information Technology Department} \\
\textit{The General Secretariat of the Council of Ministers}\\
Baghdad, Iraq \\
https://orcid.org/0009-0001-1452-6965\\
jhyondon@gmail.com}
\and
\IEEEauthorblockN{2\textsuperscript{nd}Mohamad Fadli Bin Zolkipli}
\IEEEauthorblockA{\textit{School of Computing} \\
\textit{Universiti Utara Malaysia}\\
Sintok, Malaysia \\
https://orcid.org/0000-0002-6780-8480\\
m.fadli.zolkipli@uum.edu.my}
\and
\IEEEauthorblockN{3\textsuperscript{rd}Nik Fatinah N Mohd Farid}
\IEEEauthorblockA{\textit{School of Computing}  \\
\textit{Universiti Utara Malaysia}\\
Sintok, Malaysia \\
https://orcid.org/0000-0001-5213-0016\\
nikfatinah@uum.edu.my}
\and
\IEEEauthorblockN{4\textsuperscript{th}Mohammed Alsamman, Member, IEEE}
\IEEEauthorblockA{\textit{School of Computing}  \\
\textit{Universiti Utara Malaysia}\\
Sintok, Malaysia \\
https://orcid.org/0000-0001-8929-0582\\
alsamman@uum.edu.my}
\and
\IEEEauthorblockN{5\textsuperscript{th}Zainab Ali Mohammed}
\IEEEauthorblockA{\textit{General Directorate of Urban Planning}  \\
\textit{Ministry of Construction (MCHMPW)}\\
Baghdad, Iraq  \\
https://orcid.org/0009-0009-3690-9631\\
zainab.alsalman88@gmail.com}
}

\maketitle
\IEEEpubidadjcol
\begin{abstract}In response to the escalating cyber threats, the efficiency of Cyber Threat Intelligence (CTI) data collection has become paramount in ensuring robust cybersecurity. However, existing works encounter significant challenges in preprocessing large volumes of multilingual threat data, leading to inefficiencies in real-time threat analysis. This paper presents a systematic review of current techniques aimed at enhancing CTI data collection efficiency. Additionally, it proposes a conceptual model to further advance the effectiveness of threat intelligence feeds. Following the PRISMA guidelines, the review examines relevant studies from the Scopus database, highlighting the critical role of artificial intelligence (AI) and machine learning models in optimizing CTI data preprocessing. The findings underscore the importance of AI-driven methods, particularly supervised and unsupervised learning, in significantly improving the accuracy of threat detection and event extraction, thereby strengthening cybersecurity. Furthermore, the study identifies a gap in the existing research and introduces XBC conceptual model integrating XLM-RoBERTa,  BiGRU, and CRF, specifically developed to address this gap. This paper contributes conceptually to the field by providing a detailed analysis of current CTI data collection techniques and introducing an innovative conceptual model to enhance future threat intelligence capabilities.
\end{abstract}

\begin{IEEEkeywords}
Cyber threat intelligence , Technique, Efficiency, Intelligence feeds
\end{IEEEkeywords}

\section{Introduction}

In the rapidly developing landscape of cybersecurity, the need for efficient collection, preprocessing, and analysis of cyber threat intelligence (CTI), has become a major key to protect information systems. As cyber threats grow day by day, in complexity and evolve, traditional security approaches are showing increasing weakness, leading to a need for advanced methodologies and techniques that are evidence-based approaches to threat detection and response. The integration of ML with other AI techniques into CTI systems has arisen as a powerful way to enhance the detection and mitigation of both well-known and unknown threats. The technologies that have the  ability to process huge amounts of data quickly and accurately, have changed the way that enterprises approach cyber security, making more proactive and stronger defenses \cite{Preuveneers2021140}. The role of AI especially  in optimizing CTI preprocessing, is crucial in figuring out the dynamic nature of modern cyber threats and providing a base for more powerful cybersecurity frameworks.

Despite the capabilities of these advanced technologies, important challenges keep going in optimizing and enhancing the efficiency of the collection phase within CTI. One of the major issues is the huge volume of data that generated by different threat sources, which can affect traditional data preprocessing and processing systems badly and restrict real-time analysis. Furthermore, the complexity of threat analysis in real-time, mixed with the dynamic and developing nature of cyber threats, shows real  barrier to effective CTI. The different types of threat data sources whose varying formats, structure, languages, and quality, make the collection and analysis efforts more complicated, and create inefficiency in processing, and mitigate accuracy in the threat detection. These challenges are increasing in multilingual contexts, where the need to process threat data in multiple languages adds an additional layer of complexity. Recent progress like the one by Xiang, has introduced novel methods for enhancing the efficiency of CTI collection, preprocessing, and event classification, related to the context of handling Chinese datasets \cite{Xiang2023}. However, there is still a critical need for inclusive solutions that can overcome these challenges on a wider scale.

This paper systematically identifies and evaluates current papers that aim to improve CTI feeds' collection efficiency, with a special focusing on the integration of AI and ML models within data preprocessing phase. By following the PRISMA guidelines, this review synthesizes the latest research findings from the Scopus database, providing a comprehensive understanding of the state of CTI collection methodologies. Moreover, this paper proposes the XBC conceptual model for threat intelligence event extraction within the CTI collection process, integrating  XLM-RoBERTa, BiGRU, and CRF. This model is designed to address the identified challenges by enhancing the efficiency of multilingual data processing, improving the accuracy and speed of threat detection, and ultimately contributing to the development of more effective and efficient CTI frameworks. The findings of this review are expected to support cybersecurity operations by providing a roadmap for the implementation of advanced CTI collection techniques, thereby strengthening the overall resilience of information systems against increasingly sophisticated cyber threats.

\section{Methodology}
\begin{figure}[!htb]
    \centering
    \includegraphics[width=0.95\linewidth]{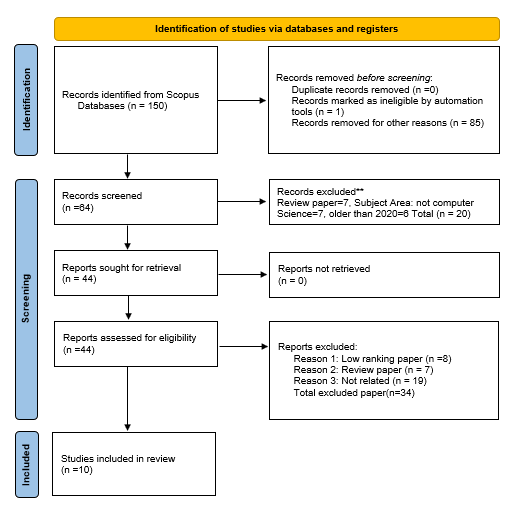}
    \caption{PRISMA flow diagram}
    \label{fig:Prisma}
\end{figure}

In this paper the PRISMA guidelines have been used, a search conducted on 27/07/2024 in Scopus database using these searching keywords: Technique, Enhance, Efficiency, “Threat Intelligence Feeds”, the first three keywords are single word, while the fourth one is multi word, to extend the result papers, alternative words are used for each keyword joined by OR operator, an AND operator was used to join different keywords, the search done within all fields because it is a very narrow area and searching only in Title, abstract and keywords gave us only 6 records before filtering. The full query is: ( ALL ( enhance OR improve OR boost OR strengthen OR augment OR upgrade OR amplify OR enrich OR elevate OR refine OR fortify OR intensify OR optimize ) AND ALL ( technique OR approach OR procedure OR strategy OR tactic OR process OR system OR model OR practice OR method OR mechanism ) AND ALL ( efficiency OR productivity OR effectiveness OR performance OR optimization OR capability OR competence OR efficacy OR proficiency OR utilization OR expediency ) AND ALL ( "Threat Data Streams" OR "Threat Intelligence Sources" OR "Security Intelligence Feeds" OR "Cyber security Intelligence Streams" OR "Malware Intelligence Feeds" OR "Threat Monitoring Feeds" OR "Cyber Threat Data" OR "Intrusion Intelligence Feeds" OR "Threat Information Feeds" OR "Attack Intelligence Feeds" OR "Network Threat Feeds" OR "Cyber Risk Intelligence Feeds" OR "Adversary Intelligence Feeds" OR "Cyber Threat Reports" OR "Security Threat Feeds" OR "Cyber Alert Feeds" OR "Threat Indicator Feeds" OR "Cyber Incident Data" OR "Cyber Threat Alerts" OR "Security Event Feeds" OR "Threat Intelligence Feeds" ) ). Figure.\ref{fig:Prisma} illustrates PRISMA phases of paper filtration and screening. The result of eligible papers is ten documents.  A result of 150 papers was achieved before any manual and automatic filtering and excluding steps. Identification and screening phases were done to these papers. An automatic filtering has been performed to exclude ineligible non-English

\begin{figure}[!htb]
    \centering
    \begin{minipage}{0.99\linewidth}
        \centering
        \includegraphics[width=\textwidth]{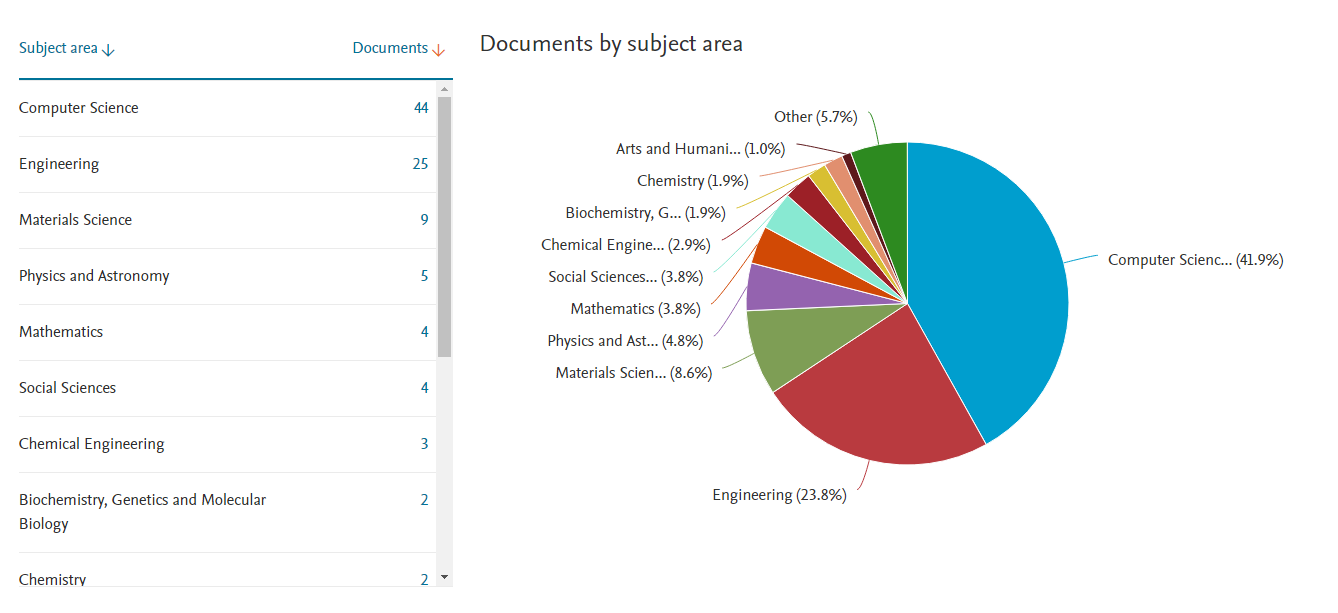}
        \caption{Document distribution by subject area}
        \label{fig:44_by_Subject_area}
    \end{minipage}

   \begin{minipage}{0.99\linewidth}
        \centering
        \includegraphics[width=\textwidth]{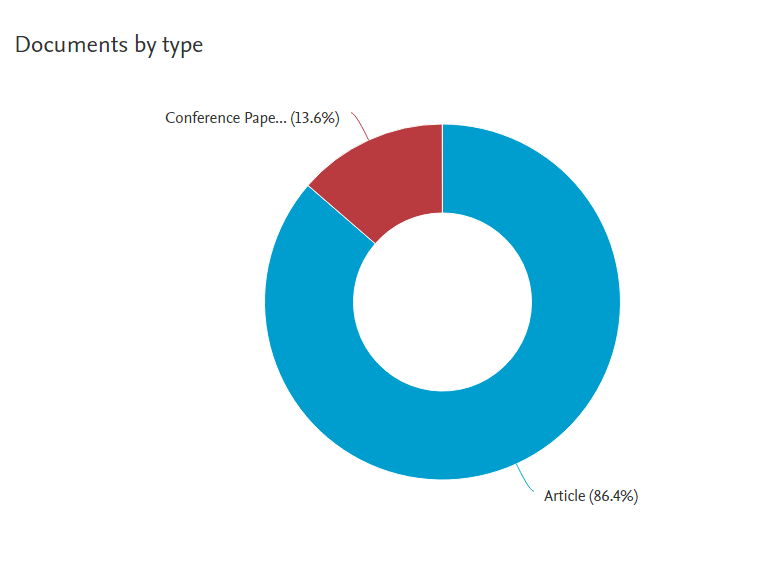}
        \caption{Document distribution by type}
        \label{fig:44_by_type}
    \end{minipage}
\hfill
    \begin{minipage}{0.99\linewidth}
        \centering
        \includegraphics[width=\textwidth]{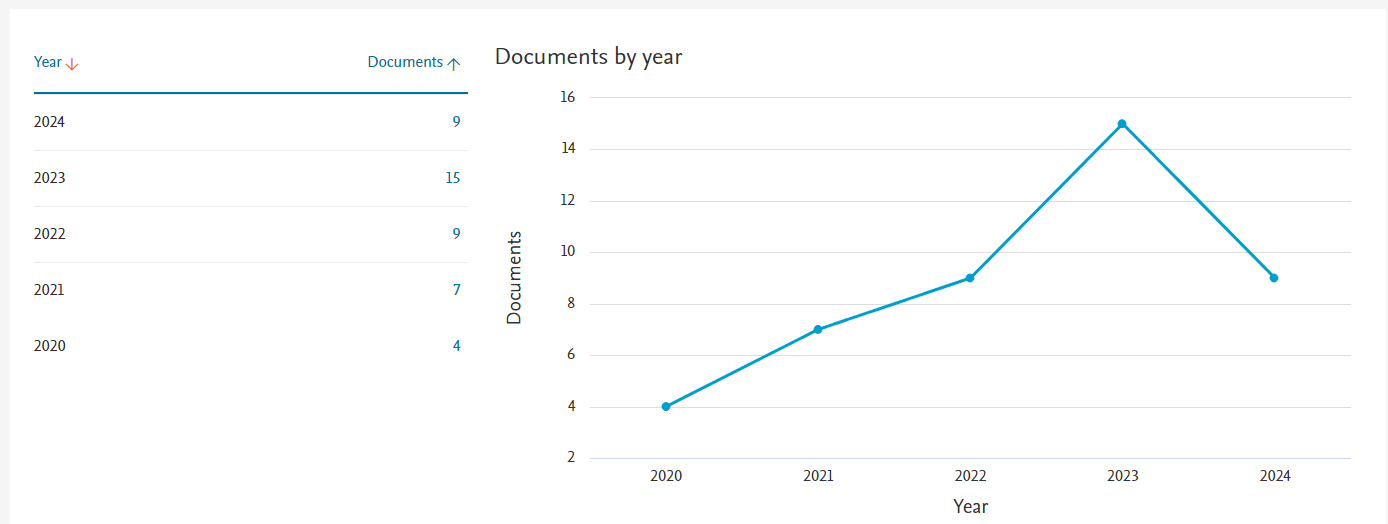}
        \caption{Document count by year}
        \label{fig:44_by_year}
    \end{minipage}
    \hfill
    \begin{minipage}{0.99\linewidth}
        \centering
        \includegraphics[width=\textwidth]{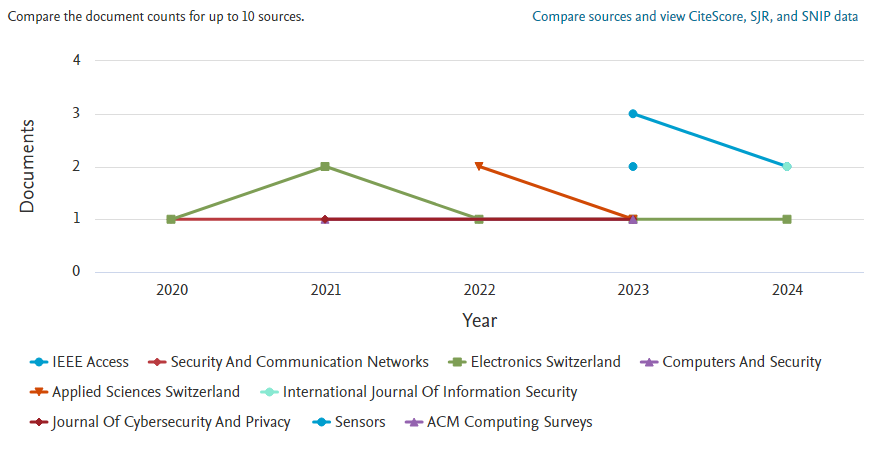}
        \caption{Timeline of documents publications by year and source}
        \label{fig:44_by_year_by_source}
    \end{minipage}
\end{figure}

1 article, and 85 articles that non accessible (not open access), 7 documents were review papers, other seven papers was not under computer science subject and lastly older than 2020 paper were six. The result for automatic filtering is forty-four papers. All papers belongs to the subject of computer science and also some of them belong to other subjects at the same time, for example 25 research lies in Engineering and 9 in Materials Science as shown in Figure.\ref{fig:44_by_Subject_area}. In the term or document types and as shown in Figure.\ref{fig:44_by_type}, 86\% of them are Article but only about 14\% are Conference paper. However all document are new released from 2020 to 2024 as mentioned before each year has different number of documents that published in, its obvious in Figure.\ref{fig:44_by_year} that 2023 has the most. On the other hand the Figure.\ref{fig:44_by_year_by_source} shows the corresponding relation between document with year and source.
Forty four papers have been assessed manually for eligibility by partially

or fully reading. All papers that not highly ranked were excluded, for instance, each of Tang\cite{Tang2023235}, Zeng\cite{Zeng2022}, Chu\cite{Chu2023}, and Czekster\cite{Czekster2022} have Q3 article, while each of Alnahari\cite{Alnahari2021}, Kumarasinghe\cite{Kumarasinghe202449}, Apruzzese\cite{Apruzzese202220}, and Sharma\cite{Sharma20238} have conference paper with low H-Index. An exclusion has been done for all paper that was based on interviews and survey (Vevera\cite{Vevera202213}, Varga\cite{Varga2021}, and Stojkovski\cite{Stojkovski2021385}). Many papers have found as a review paper while reading and were excluded, for example: Ferdous\cite{Ferdous2023121118}, van Haastrecht\cite{van_Haastrecht2021}, Browne\cite{Browne20242911}, Alzahrani\cite{Alzahrani2024}, Ispahany\cite{Ispahany202468785}, de Melo\cite{de_Melo_e_Silva2020},  and Xin\cite{Xin20222835}.
A group of sixteen papers were excluded due to non-relevancy as follows: Asiri\cite{Asiri2023}, in his article, investigated coping strategies for mobile malware, which is unrelated to threat intelligence data streams. Burke\cite{Burke2024} advocated self-evaluation for cybersecurity, but this does not contribute to enhancing intelligence data processing. Smyrlis\cite{Smyrlis20241821} discussed risk assessment in healthcare, which does not address the efficiency of intelligence gathering. Gulatas\cite{Gulatas202333584} addressed malware in IoT environments without focusing on improving threat data handling. Rantos\cite{Rantos2020} discussed interoperability challenges in information sharing rather than enhancing the efficiency of intelligence data collection. Lee\cite{Lee2021} examined the effectiveness of security policies, which is not relevant to the efficiency of threat intelligence gathering. Almomani\cite{Almomani2023} focused on ransomware detection, not on optimizing intelligence data efficiency. Preuveneers\cite{Preuveneers2020} proposed a security framework that enhances the reliability of intelligence sharing, but it lacks a focus on the efficiency of data processing. Bouramdane\cite{Bouramdane2023662} discussed cyberattacks on smart grids, which does not relate to the optimization of intelligence data collection. Crotty\cite{Crotty2022} focused on cyber risk assessment, not on the efficiency of intelligence data streams. Jolles\cite{Jollès2023140} examined the dynamics of information sharing, without addressing data collection efficiency. Jorquera\cite{Jorquera_Valero2023} discussed security and trust in 5G networks, without focusing on enhancing threat data processing. Gong\cite{Gong20211} proposed a framework for incident response, which does not target the efficiency of intelligence data. Sakellariou\cite{Sakellariou2022} offered a reference model for CTI, but did not specifically address the efficiency of data collection. Coscia\cite{Coscia2024} focused on the generation of rulesets for NIDPS, not on improving intelligence data efficiency. Chatziamanetoglou\cite{Chatziamanetoglou2023} discussed the use of blockchain for intelligence sharing, but lacked a focus on the efficiency of data streams. 

\section{Literature Rivew}
Zoppi, in his article \cite{Zoppi2023}, compares 47 supervised, unsupervised, and meta-learning algorithms for intrusion detection across 11 datasets. He highlights that unsupervised meta-learning algorithms, particularly boosting ensembles, outperform others in detecting unknown cyber threats, making them suitable for enhancing cyber threat intelligence collection efficiency. Preuveneers \cite{Preuveneers2021140} presents a framework for improving cyber threat intelligence collection by sharing machine learning models. His work enhances detection accuracy and protects against adversarial attacks, emphasizing the importance of encryption and collaborative improvement. He also proposes future work on automated annotation and expanded use cases.

Ahmed \cite{Ahmed2024} explores the integration of honeypot data with machine learning techniques to enhance cyber threat detection in IoT-based smart cities. His study addresses the growing vulnerabilities in IoT devices within smart cities by proposing a framework that improves attack detection and mitigation through the analysis of real-world cyber-attack datasets. The findings demonstrate significant improvements in security when incorporating honeypot data into IoT security frameworks, filling a crucial gap in existing research.

Ali \cite{Ali20201} investigates an approach to enhance malware detection efficiency by leveraging N-grams and machine learning. He proposes a dynamic analysis technique using AI-based sandboxing to extract features, achieving high classification accuracy, particularly with Logistic Regression.

Ilca \cite{Ilca2023} presents a comprehensive, open-source cybersecurity solution tailored for SMEs. His solution leverages advanced malware detection techniques, machine learning, and real-time threat intelligence to enhance the cyber-resilience of organizations with limited resources. The system's scalability and affordability make it an effective defense against emerging threats, ensuring proactive malware detection and robust incident response capabilities.

Cherqi \cite{Cherqi202384440} explores the improvement of efficiency in cyber threat intelligence collection by employing contrastive learning methods to enhance data representation and threat detection accuracy. His work emphasizes robust model performance in identifying evolving cyber threats.

Kim \cite{Kim2022} evaluates methods to enhance TTP classification from cyber threat intelligence (CTI) datasets, focusing on addressing class imbalance using oversampling techniques like SMOTE and EDA. The improved classification accuracy in his study indicates the effectiveness of these augmentation strategies, which are crucial for advancing automated CTI processing.
Bazlur \cite{Bazlur_Rashid2022} introduces an anomaly detection approach that utilizes feature selection to enhance performance. His method reduces dataset features, thereby improving detection accuracy and computational efficiency in cybersecurity. Results across various datasets indicate significant improvements in true positive rates and reduced false positive rates compared to existing methods. This approach proves beneficial in efficiently handling large-scale cybersecurity data, making it a valuable tool in enhancing threat detection.

Mohanty \cite{Mohanty202450578} proposes a hybrid approach that combines filter and wrapper feature selection methods to enhance the efficiency of detecting malicious URLs in IoT devices. His approach achieves a 98.3\% accuracy and is optimized for resource-constrained environments.
Xiang \cite{Xiang2023} presents a novel APT event extraction method using the BERT-BiGRU-CRF model to enhance APT attack detection. By defining an APT event schema and constructing a Chinese dataset, his model significantly improves extraction performance, thereby aiding more effective cyber threat intelligence analysis.

\section{Discussion}
This paper has analyzed the related articles intensively and analyzed their outcomes which can be grouped into three categories. The first group includes works by Xiang\cite{Xiang2023}, Mohanty\cite{Mohanty202450578}, Bazlur\cite{Bazlur_Rashid2022}, Cherqi\cite{Cherqi202384440}, Ali\cite{Ali20201}, and Ahmed\cite{Ahmed2024}, all of them focus on enhancing detection techniques in various areas of cyber security, such as APT event extraction, URL detection, anomaly detection, and IoT threat detection. The second group, consisting of Kim\cite{Kim2022} and Zoppi\cite{Zoppi2023}, centers on improving classification and detection models, particularly through TTP classification and meta-learning for unknown threats. The final group includes Ilca\cite{Ilca2023} and Preuveneers\cite{Preuveneers2021140}, who focus on security improvements within specific contexts, such as SMEs and general system security, through enhanced malware detection and model robustness. Table.\ref{tab:comp1} demonstrates the literature summary for these related paper.
\setlength{\tabcolsep}{3pt} 

\begin{table*}[h]

  \centering
  \renewcommand{\arraystretch}{0.8}
  \setlength{\tabcolsep}{4pt}
  \caption{Literature review summary table}
   \resizebox{1\textwidth}{!}{
   \scriptsize
    
    \begin{tabularx}{\textwidth}{cccccp{4cm}p{4cm}p{4cm}}
    No & \multicolumn{1}{c}{Quar.} & \multicolumn{1}{c}{\shortstack{Cited \\ by}} & \multicolumn{1}{c}{Year}& \multicolumn{1}{c}{\shortstack{Main\\Author}} & \multicolumn{1}{c}{Outcomes} & \multicolumn{1}{c}{Limitations} & \multicolumn{1}{c}{\shortstack{ \\Future Work}} \\
    \hline
    1     & Q2    & 6     & 2023  & Xiang & Enhanced APT event extraction improves CTI efficiency. & Limited dataset size, focuses only on Chinese & Expand dataset, apply model to multilingual contexts. \\
    2     & Q1    & 0     & 2024  & Mohanty & Hybrid approach improves URL detection accuracy, efficiency. & Limited real-time testing and computational comparisons explored. & Optimize feature selection, improve real-time detection \\
    3     & Q2    & 21    & 2022  & Bazlur & Enhanced anomaly detection via feature selection techniques & Limited evaluation with newer datasets and algorithms. & Explore optimizers and techniques for higher efficiency. \\
    4     & Q2    & 3     & 2022  & Kim   & Improved TTP classification using oversampling techniques significantly & Prone to generalization errors; needs quality data. & Focus on quality data, model optimization, embedding. \\
    5     & Q1    & 0     & 2023  & Cherqi & Improved threat detection accuracy via contrastive learning. & Limited generalizability and dataset dependency for effectiveness. & Exploration of diverse datasets and real-world applications. \\
    6     & Q2    & 11    & 2023  & Ilca  & Enhanced malware detection and response for SMEs. & Expand system scalability and improve container integration. & Expand system scalability and improve container integration. \\
    7     & Q2    & 38    & 2020  & Ali   & High detection accuracy using N-grams and machine learning. & Limited dataset size and feature scope tested. & Expand dataset, explore more features, implement deep learning. \\
    8     & Q2    & 0     & 2024  & Ahmed & Enhanced IoT threat detection via honeypot data. & Limited dataset availability and specific IoT focus. & Extend honeypot use, integrate with IDS systems. \\
    9     & Q2    & 32    & 2021  & Preuveneers & Improved detection, model sharing, encryption, robustness. & Adversarial vulnerabilities, computational overhead, limited scope. & Automated annotation, expanded use cases, explainability. \\
    10    & Q1    & 22    & 2023  & Zoppi & Unsupervised meta-learning enhances unknown threat detection. & Heterogeneous datasets, suboptimal hyperparameters, limited real-world applicability. & Refine models, expand datasets, validate with real-world data. \\
    \end{tabularx}
    }
   
    \label{tab:comp1}
\end{table*}

\setlength{\tabcolsep}{10pt} 
\begin{table*}[!h]

  \centering
  \renewcommand{\arraystretch}{1}
   \caption{Achieved accuracy for used algorithm vs articles, article number taken from Table \ref{tab:comp1}}
   \resizebox{\textwidth}{!}{
   \centering
    \begin{tabularx}{\textwidth}{lcccccccccc}
    Algorithm/Article & \multicolumn{1}{l}{1} & \multicolumn{1}{l}{2} & \multicolumn{1}{l}{3} & \multicolumn{1}{l}{4} & \multicolumn{1}{l}{5} & \multicolumn{1}{l}{6} & \multicolumn{1}{l}{7} & \multicolumn{1}{l}{8} & \multicolumn{1}{l}{9} & \multicolumn{1}{l}{10} \\
    \hline
    ADABoost& -     & 96.5\% & -     & -     & -     & -     & -     & -     & -     & 92.1\% \\
    Autoencoder& -     & -     & -     & -     & -     & -     & -     & -     & 99.58\% & - \\
    AutoGluon& -     & -     & -     & -     & -     & -     & -     & -     & -     & 90.1\% \\
    Bagging & -     & 97.4\% & -     & -     & -     & -     & -     & -     & -     &  \\
    BERT  & 58\%  & -     & -     & -     & -     & -     & -     & -     & -     & - \\
    BERT-BiGRU-CRF& 70.1\% & -     & -     & -     & -     & -     & -     & -     & -     & - \\
    BiGRU-CRF & 52\%  & -     & -     & -     & -     & -     & -     & -     & -     & - \\
    CCFSRFG & -     & -     & 98.38\% & -     & -     & -     & -     & -     & -     & - \\
    COF   & -     & -     & -     & -     & -     & -     & -     & -     & -     & 87.4\% \\
    CL+GAN-BERT & -     & -     & -     & -     & 81\%  & -     & -     & -     & -     & - \\
    DT & -     & -     & -     & -     & -     & -     & 78.79\% & 99.89\% & 99.9996\% & 93.5\% \\
    FastABOD & -     & -     & -     & -     & -     & -     & -     & -     & -     & 85.3\% \\
    FastAI & -     & -     & -     & -     & -     & -     & -     & -     & -     & 88.3\% \\
    G-Means & -     & -     & -     & -     & -     & -     & -     & -     & -     & 85.9\% \\
    GB & -     & 98\%  & -     & -     & -     & -     & -     & -     & -     & 93.5\% \\
    HBOS  & -     & -     & -     & -     & -     & -     & -     & -     & -     & 87.5\% \\
    iForest & -     & -     & -     & -     & -     & -     & -     & -     & -     & 87.6\% \\
    K-Means & -     & -     & -     & -     & -     & -     & -     & -     & -     & 87.1\% \\
    KNN   & -     & 94.4\% & -     & -     & -     & -     & -     & 99.2\% & -     & 93.6\% \\
    LDCOF & -     & -     & -     & -     & -     & -     & -     & -     & -     & 85.4\% \\
    (LDA) & -     & -     & -     & -     & -     & -     & -     & -     & -     & 86.2\% \\
    LOF   & -     & -     & -     & -     & -     & -     & -     & -     & -     & 83.8\% \\
    LG & -     & -     & -     & 94.8\% & -     & -     & 84.5\% & -     & -     & 80.8\% \\
    LSTM  & -     & -     & -     & -     & -     & -     & -     & 98.21\% & -     & - \\
    MIG+GA & -     & 98.3\% & -     & -     & -     & -     & -     & -     & -     & - \\
    MLP   & -     & -     & -     & 95\%  & -     & -     & -     & -     & 99.97\% & - \\
    Naive Bayes & -     & -     & 94\%  & 83.6\% & -     & -     & 82.83\% & 95.15\% & -     & 80.3\% \\
    ODIN  & -     & -     & -     & -     & -     & -     & -     & -     & -     & 87.6\% \\
    One-Class SVM & -     & -     & -     & -     & -     & 78.79\% & 83.43\% & -     & 85.73\% & 87.1\% \\
    Py-Custom & -     & -     & -     & -     & -     & -     & -     & -     & -     & 89.5\% \\
    Random Forest & -     & -     & -     & -     & -     & -     & 79.8\% & -     & 99.9998\% & 94.5\% \\
    SDO   & -     & -     & -     & -     & -     & -     & -     & -     & -     & 86.7\% \\
     SNN  & -     & -     & -     & -     & -     & -     & -     & 98.96\% & -     & - \\
    SOM   & -     & -     & -     & -     & -     & -     & -     & -     & -     & 85.3\% \\
    (SVMs) & -     & -     & -     & -     & -     & -     & -     & -     & -     & 82\% \\
    TabNet & -     & -     & -     & -     & -     & -     & -     & -     & -     & 87.9\% \\
    XGBoost & -     & 98.3\% & -     & -     & -     & -     & -     & -     & -     & 94.8\% \\
    \end{tabularx}
    }

    \label{tab:my_label}
  
  \end{table*}

\begin{figure}[b]
    \centering
    \includegraphics[width=0.6\linewidth]{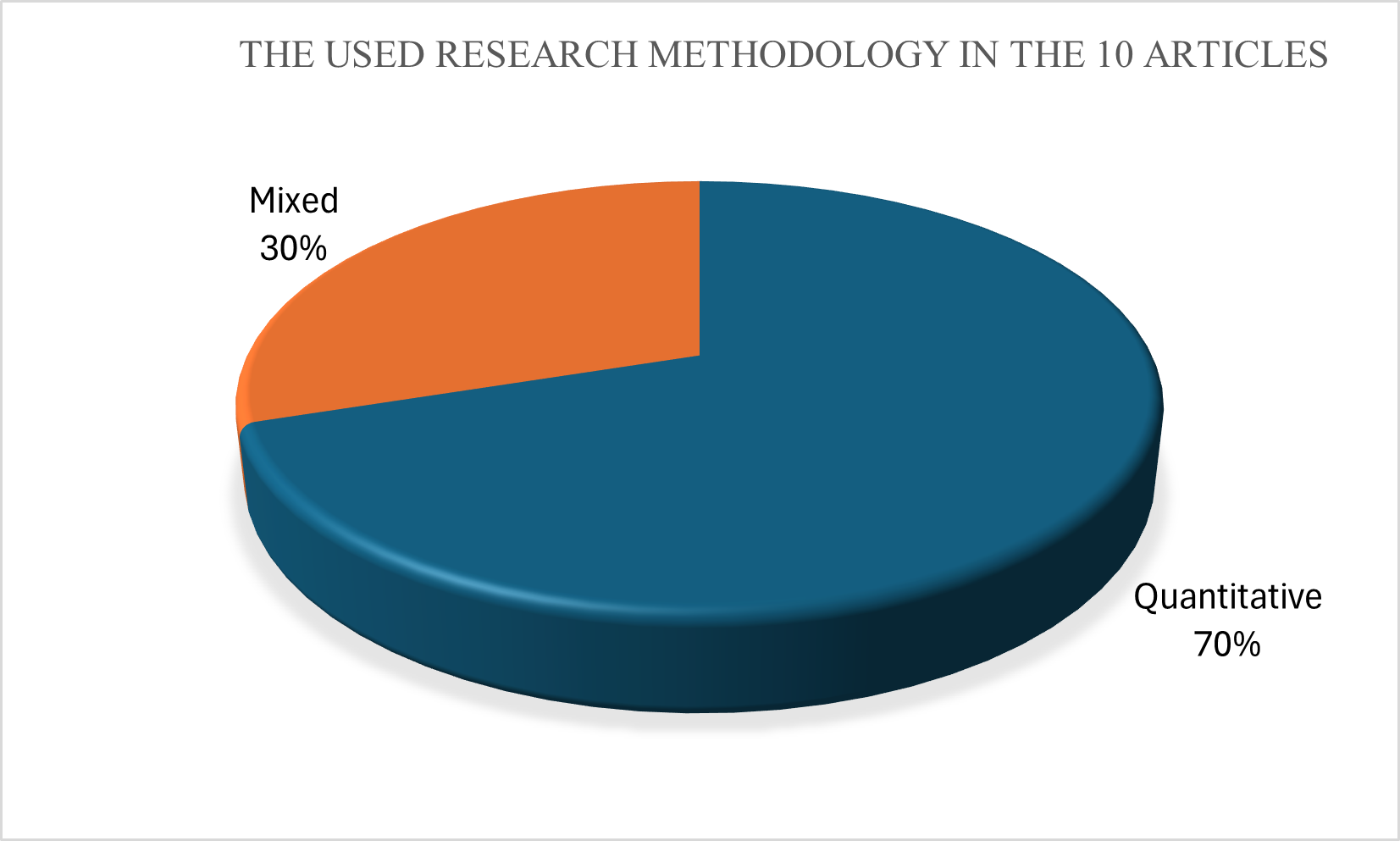}
    \caption{Proportional usage of research methodologies in the related papers}
    \label{fig:10_the_Used_methodology}
\end{figure}

Figure.\ref{fig:10_the_Used_methodology} shows that 70\% of these articles followed quantitative research methodology while 30\% of them had mixed research methodology.

This paper showed AI algorithms have major key influence on enhancing efficiency in CTI feeds collection and event extraction. A wide range of algorithms have been used giving a variety of solutions and methods whose evaluation metrics Accuracy, Precision, and Recall and F1-Score.

\begin{figure}[b]
    \centering
    \includegraphics[width=1\linewidth]{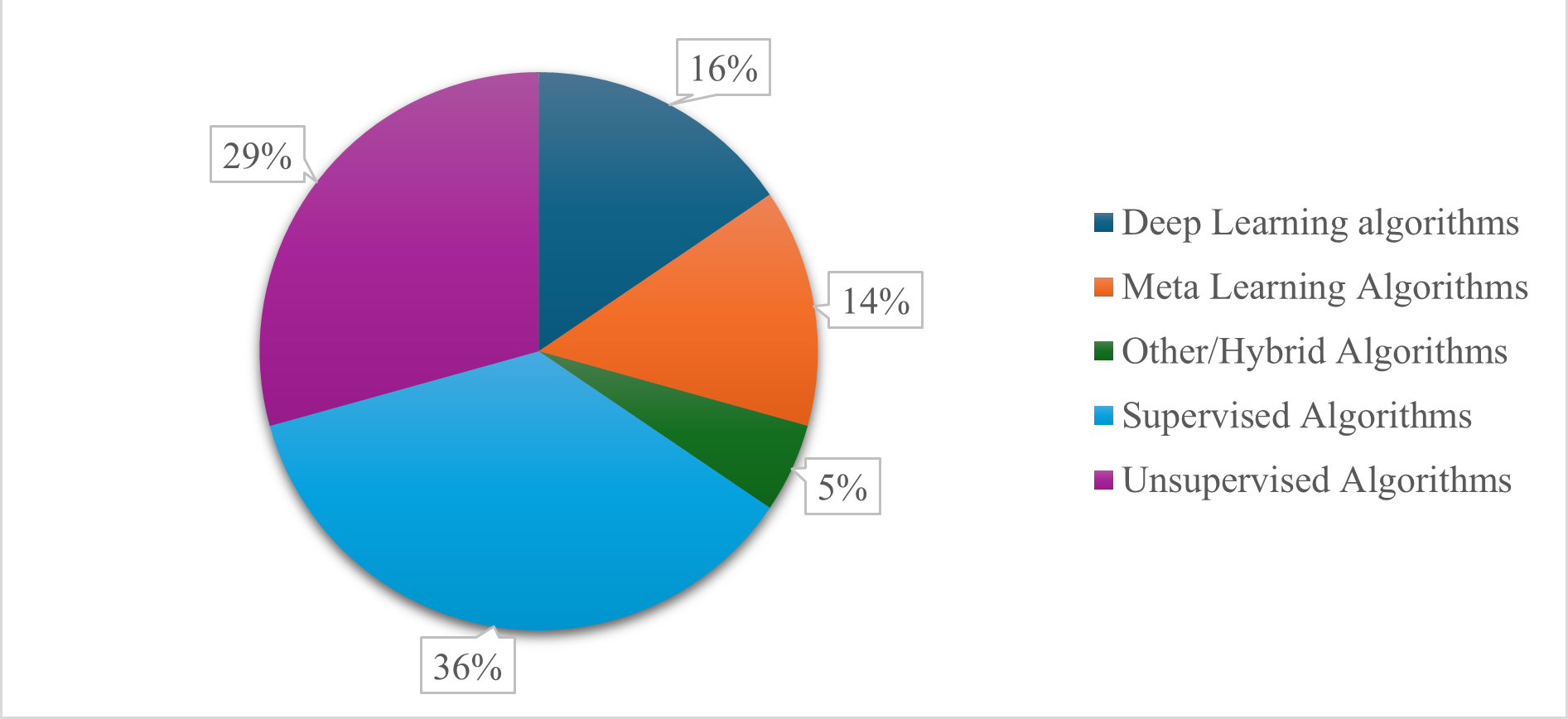}
    \caption{Proportional distribution of algorithm types in related papers}
    \label{fig:10_algorithims_types_vs_articles}
\end{figure}

This study focused on accuracy and made matrix table showing an crucial overview for these important article vs the used algorithm and methods in their work and the maximum achieved value of classification accuracy in collection phase of CTI as in Table.\ref{tab:my_label}. Figure.\ref{fig:10_the_Used_methodology} shows how frequently these algorithm has been used in these 10 articles.Furthermore these algorithms are classified into: supervised, unsupervised, meta learning, Deep learning and Other/Hybrid, Figure.\ref{fig:10_algorithims_types_vs_articles} demonstrates these proportions.
The future work outlined in the articles can be grouped into three categories. First, expanding datasets and improving model applicability is a common theme, as seen in the works of Xiang\cite{Xiang2023}, Ali\cite{Ali20201}, Cherqi\cite{Cherqi202384440}, and Zoppi\cite{Zoppi2023}, all of them propose broadening the data used to test and refine their models to enhance generalizability and effectiveness in diverse contexts. The second category focuses on optimizing and refining models and techniques,  Mohanty\cite{Mohanty202450578}, Bazlur\cite{Bazlur_Rashid2022}, Kim\cite{Kim2022}, and Zoppi\cite{Zoppi2023} suggesting improvements in feature selection, real-time detection, and model efficiency to address current limitations in computational performance and accuracy. Lastly, extending the application scope and integration capabilities is emphasized: Ahmed\cite{Ahmed2024}, Ilca\cite{Ilca2023}, and Preuveneers\cite{Preuveneers2020}, who propose integrating their models with additional systems, such as IDS, and exploring more use cases to increase the practicality and robustness of their solutions in real-world scenarios. These future directions reflect a shared goal of advancing cyber security through broader data use, enhanced model optimization, and wider application integration.
In terms of limitations, several studies, including those by Ali\cite{Ali20201}, and Ahmed\cite{Ahmed2024}, are constrained by limited dataset sizes, impacting the generalizability and robustness of their findings. Another group, including Mohanty\cite{Mohanty202450578}, Bazlur\cite{Bazlur_Rashid2022}, Kim\cite{Kim2022}, and Zoppi\cite{Zoppi2023}, faces challenges related to computational efficiency, real-time testing, and optimization, which limit their practical applicability and effectiveness. Lastly, works by Cherqi\cite{Cherqi202384440}, Ilca\cite{Ilca2023}, and Preuveneers\cite{Preuveneers2021140} highlight issues of scalability and dataset dependency, where the effectiveness of their models is heavily reliant on the quality and diversity of data, along with concerns about computational overhead and adversarial vulnerabilities. On the other hand, Xiang\cite{Xiang2023}'s work showed a limitation in his model of dealing with only

\begin{figure*}[h]
\centering
    \includegraphics[width=1\textwidth]{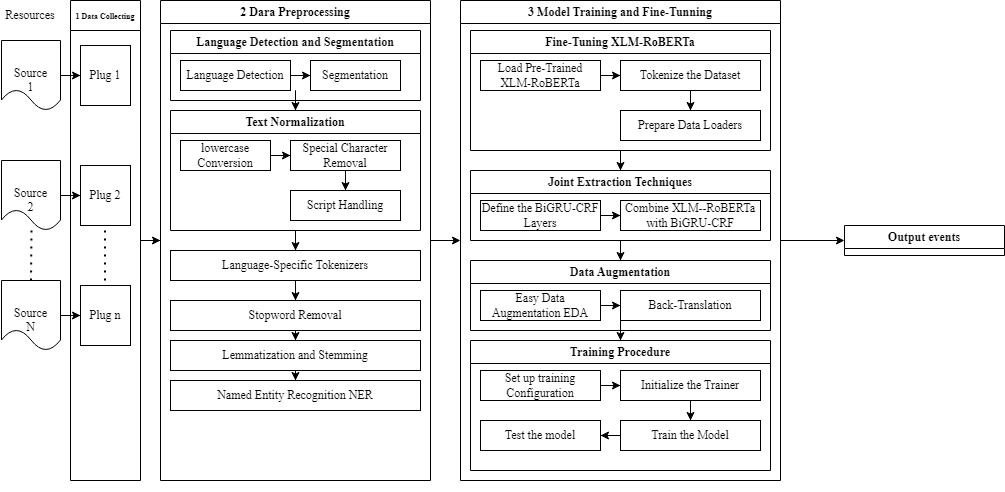}
    \caption{XBC Model's flow chart}
    \label{fig:XBC}
\end{figure*}
Chinese language that extracted from context different resources of CTI feeds, that motivated us to fill his gap of monolingual and overcome it by our proposed conceptual model XBC.
\section{Proposed Work}
The article by Acs\cite{Acs2023}. on morphosyntactic probing of multilingual XLM-RoBERTa model provides significant insights into the strengths of this model in handling multilingual data. Acs et al. demonstrated that XLMRoBERTa learn strong multilingual representations, outperforming traditional models in various linguistic tasks. 
This finding supports the potential of XLMRoBERTa to address the monolingual limitations of Xiang et al.’s model.
The proposed conceptual model XBC outlines a comprehensive methodology for extracting events from multilingual text sources using advanced natural language processing techniques. The methodology is divided into three primary phases: Data Collection, Data Preprocessing, and Model Training and Fine-Tuning. Initially, data is collected from cybersecurity-focused Facebook groups, Hack Forums, and CERT-IN using automated web scraping techniques and APIs. Specifically, APIs and plug-ins (Facebook’s Graph API, Scrapy, and Requests) are used to gather data from these sources, and store it in JSON format.

The language of each data entry is automatically identified and categorized to handle multilingual data effectively, starting with the Polyglot language detection library to determine the language of each text entry. The \textbf{Pandas} library applies segmentation to the collected context and divides it into different subsets based on the detected language. The  \textbf{JSON} format will be used for storing the output, as JSON is simple format for both humen and AI.

In the next step, text normalization takes place, starting with converting all context to lowercase, then removing special characters, and script handling using  \textbf{Pandas, Regular Expressions, and Unicode libraries.  SpaCy} will be used to break down the text into individual words or tokens that relate to the specific language that is under processing. For example, in English, if the sentence is “This paper is a conceptual” stopwords like ‘This,’ ‘is,’ and ‘a’ will be removed, leaving only meaningful words like “paper” and “conceptual.” \textbf{SpaCy} will also be used for lemmatization, where reducing words to their origin base or dictionary form \textbf{(lemma)}, for instance "running" lemmatize to "run"), then the Named Entity Recognition (NER) process will take place; which is an NLP task that identifies and classifies key information within text into categories such as person names, organizations, locations, dates, products, and events. NER will extract structured information from unstructured input data that has already passed the lemmatization process, applying the information to organizing, analyzing, and retrieving relevant details. In this multilingual CTI model \textbf{(XBC)}, the NER process focuses on crucial entities, such as threat actors, IP addresses, attackes techniques, and events, enhancing the accuracy and reliability of the analysis from diverse sources.

Later on \textbf{NLTK} (Natural Language Toolkit) will be used for removing and stemming stopwords , where to cut words down to their root form, often by simply removing suffixes, if we use the same example of the word "running," it will be stemming to "runn" after cutting the suffex "ing" without considering the context.
After this step the output of the preprocessing phase is stored in \textbf{Parquet} format.

Model training and fine-tuning are crucial steps in developing an effective and reliable multilingual CTI model. XLM-RoBERTa is employed to use its superior capabilities in dealing with multilingual contexts, escpecially for classifying tactics, techniques, and procedures (TTPs) within multilingual CTI feeds. The integration process starts with fine-tuning XLM-RoBERTa on a multilingual CTI dataset, adapting it to the required TTP classification. The XBC model architecture is enhanced by interating XLM-RoBERTa with BiGRU and CRF layers. This combination supports and gather the contextual embeddings from XLM-RoBERTa with the sequence modeling skills of BiGRU, alongside the structured prediction abilities of CRF, by using this way of applying a joint extraction mechanism. The architecture is further developed into a unified model that combines these layers with high performance over different languages.

Data collection phase is very crucial to the research, focusing on gathering from multilingual CTI dataset. Data is sourced from cyber security-focused Facebook groups, online hacker forums , and CERT-IN, using automated web scraping tools, plugins, and APIs. This mixed data collection approach ensures the inclusion of threat intelligence in various languages and from wider ragne of resources.

Preprocessing the collected data is a essential step, eo enshure that  the data is clean, consistent, and ready for model training. Preprocessing includes several sequenced steps: language detection, text normalization, tokenization, stopword removal, lemmatization, stemming, and named entity recognition (NER). These steps standardize and unify the data, reducing noise and making sure of consistency, which is importante for the model effective performance.

Following preprocessing defined step, the research design involves training and fine-tuning the integrated combination of XLM-RoBERTa with BiGRU and CRF model on the preprocessed dataset. The training procedure starts with setting up the training configuration then applying joint extraction techniquesfollowed by implementing data augmentation methods such as synonym replacement and back-translation. The training process is carefully monitored, and the model's performance is evaluated using validation datasets to ensure its effectiveness in identifying and analyzing cyber threats across different languages.

The research design is discribed with a block diagram that visually represents the methodology, highlighting the flow from data collection to model training. The entire model is designed to ensure that the model performance is robust, reliable, accurate, and capable of handling the chalange of multilingual CTI data, resulting in enhancing the overall capability of cyber threat intelligence systems. Figure \ref{fig:XBC} illustrates the flowchart of the proposed conceptual model XBC.

\section{Conclusion}
This paper examined key methodologies and advancements in enhancing Cyber Threat Intelligence (CTI) collection efficiency. By following PRISMA guidelines, ten significant studies were identified and analyzed, highlighting the crucial role of artificial intelligence (AI) and machine learning (ML) in optimizing CTI processes.

Despite progress, challenges remain, particularly in managing large data volumes, real-time threat analysis, and the heterogeneity of data sources. AI-driven approaches have improved CTI efficiency, but more sophisticated methods are needed to address these issues, especially in multilingual environments.

The review identifies a gap in existing research, particularly in processing multilingual threat data. To address this, the authors proposed the XBC conceptual model, aimed at enhancing efficiency through multilingual preprocessing techniques.

In conclusion, advancing CTI methodologies is vital for improving cybersecurity in the face of increasingly sophisticated threats.

\bibliographystyle{IEEEtran}
\bibliography{ref}

\end{document}